\documentclass[%
 amsmath,amssymb,
preprint,%
 reprint,%
]{revtex4-1}

\usepackage{graphicx}%
\usepackage{dcolumn}%
\usepackage{bm}%

\usepackage[utf8]{inputenc}
\usepackage[T1]{fontenc}
\usepackage{mathptmx}

\begin{document}

\preprint{AIP/123-QED}

\title[AVSML]{Machine Learning meets Quantum Foundations: A Brief Survey}

\author{Kishor Bharti}
\email{e0016779@u.nus.edu}
\affiliation{Centre for Quantum Technologies, National University of Singapore, 3 Science Drive 2, Singapore 117543}

\author{Tobias Haug}
\affiliation{Centre for Quantum Technologies, National University of Singapore, 3 Science Drive 2, Singapore 117543}

\author{Vlatko Vedral}
\affiliation{Centre for Quantum Technologies, National University of Singapore, 3 Science Drive 2, Singapore 117543} 
\affiliation{Clarendon Laboratory, University of Oxford, Parks Road, Oxford OX1 3PU, United Kingdom }

\author{Leong-Chuan Kwek}
\affiliation{Centre for Quantum Technologies, National University of Singapore, 3 Science Drive 2, Singapore 117543}
\affiliation{MajuLab, CNRS-UNS-NUS-NTU International Joint Research Unit, UMI 3654, Singapore}
\affiliation{National Institute of Education,
Nanyang Technological University, 1 Nanyang Walk, Singapore 637616}
\affiliation{School of Electrical and Electronic Engineering
Block S2.1, 50 Nanyang Avenue, 
Singapore 639798 }

\date{\today}

\begin{abstract}
\begin{quotation}
The goal of machine learning is to facilitate a computer to execute a specific task without explicit instruction by an external party. Quantum foundations seeks to explain the conceptual and mathematical edifice of quantum theory. Recently, ideas from machine learning have successfully been applied to different problems in quantum foundations. Here, we compile the representative works done so far at the interface of machine learning and quantum foundations.  We conclude the survey with potential future directions.
\end{quotation}
\end{abstract}

\maketitle

\tableofcontents

\section{\label{sec:Intro}Introduction}

The rise of machine learning in recent times has remarkably transformed
science and society. The goal of machine learning is to get computers
to act without being explicitly programmed \cite{shalev2014understanding,goodfellow2016deep}. Some of the typical applications
of machine learning are self-driving cars, efficient web search, improved
speech recognition, enhanced understanding of the human genome and
online fraud detection. This viral spread in interest has exploded
to various areas of science and engineering, in part due to the hope
that artificial intelligence may supplement human intelligence to
understand some of the deep problems in science.

The techniques from machine-learning have been used for automated theorem
proving, drug discovery and predicting the 3-D structure of proteins
based on its genetic sequence~\cite{alquraishi2019alphafold,bridge2014machine, lavecchia2015machine}. In physics, techniques from machine
learning have been applied for many important avenues \cite{arsenault2015machine,zhang2017quantum,carrasquilla2017machine,van2017learning,deng2017machine,wang2016discovering,broecker2017machine,ch2017machine,zhang2017machine,wetzel2017unsupervised,hu2017discovering,yoshioka2018learning,torlai2016learning,huang2019near,aoki2016restricted,you2018machine,pasquato2016detecting,hezaveh2017fast,biswas2013application,abbott2016observation,kalinin2015big,schoenholz2016structural,liu2017self,huang2017accelerated,torlai2018neural,chen2018equivalence,huang2017neural,schindler2017probing,haug2019engineering,cai2018approximating,broecker2017quantum,nomura2017restricted,biamonte2017quantum,haug2020classifying}
including the study of black hole detection \cite{abbott2016observation}, topological codes \cite{torlai2017neural}, phase
transition \cite{hu2017discovering}, glassy dynamics\cite{schoenholz2016structural}, gravitational lenses\cite{hezaveh2017fast}, Monte Carlo simulation \cite{huang2017accelerated, liu2017self},
wave analysis \cite{biswas2013application}, quantum state preparation \cite{bukov2018reinforcement1, bukov2018reinforcement2}, anti-de Sitter/conformal
field theory (AdS/CFT) correspondence \cite{hashimoto2018deep}  and characterizing the landscape
of string theories \cite{carifio2017machine}. Vice versa, the methods from physics have
also transformed the field of machine learning both at the foundational
and practical front \cite{lin2017does, cichocki2014tensor}. For a comprehensive review on machine learning for
physics, refer to Carleo \textit{et al}~\cite{carleo2019machine} and references therein. For a thorough review on machine learning and artificial intelligence in the quantum domain, refer to Dunjko \textit{et al} \cite{dunjko2016quantum} or Benedetti \textit{et al} \cite{benedetti2019parameterized}.

Philosophies in science can, in general, be delineated from the study
of the science itself. Yet, in physics, the study of quantum foundations
has essentially sprouted an enormously successful area called quantum
information science. Quantum foundations tells us about the mathematical as well as conceptual understanding of quantum theory. Ironically, this area
has potentially provided the seeds for future computation and communication,
without at the moment reaching a consensus among all the physicists
regarding what quantum theory tells us about the nature of reality
\cite{hardy2010physics}.

In recent years, techniques from machine learning have been used to
solve some of the analytically/numerically complex problems in quantum
foundations. In particular, the methods from reinforcement learning
and supervised learning have been used for determination of the maximum
quantum violation of various Bell inequalities, the classification of experimental
statistics in local/nonlocal sets, training AI for playing Bell nonlocal
games, using hidden neurons as hidden variables for completion of quantum
theory, and machine learning-assisted state classification \cite{krivachy2019neural, canabarro2019machine, bharti2019teach, deng2018machine}.

Machine learning also attempts to mimic human reasoning leading to
the almost remote possibility of machine-assisted scientific discovery
\cite{iten2020discovering}. Can machine learning do the same with
the foundations of quantum theory? At a deeper level, machine learning
or artificial intelligence, presumably with some form of quantum computation,
may capture somehow the essence of Bell nonlocality
and contextuality. Of course, such speculation belies the
fact that human abstraction and reasoning could be far more complicated
than the capabilities of machines.

In this brief survey, we compile some of the representative works done so far at the interface
of quantum foundations and machine learning. The survey includes eight
sections excluding the introduction (section \ref{sec:Intro}). In
section \ref{sec:ML}, we discuss the basics of machine learning.
Section \ref{sec:Foundation} contains a brief introduction to quantum
foundations. In sections \ref{sec:Oracle} to \ref{sec:Interpretations},
we discuss various applications of machine learning in quantum foundations. There is a rich catalogue of works, which we could not incorporate in detail in sections \ref{sec:Oracle} to \ref{sec:Interpretations}, but we find them worth mentioning. We include such works briefly in section \ref{sec:more}. Finally, we conclude in section \ref{sec:Conclusion} with open questions and some speculations.

\section{\label{sec:ML} Machine Learning}

Machine learning is a branch of artificial intelligence which involves
learning from data \cite{goodfellow2016deep, shalev2014understanding}. The purpose of machine
learning is to facilitate a computer to achieve a specific task without
explicit instruction by an external party. According to Mitchel (1997) \cite{mitchell1997machine}
``\textit{A computer program is said to learn from experience $E$
with respect to some class of tasks $T$ and performance measure $P$,
if the performance at tasks in $T$, as measured by $P$, improves
with $E$.}'' Note that the meaning of the word ``task'' doesn't
involve the process of learning. For instance, if we are programming
a robot to play Go, playing Go is the task. Some of the examples of
machine learning tasks are following.
\begin{itemize}
\item Classification: In classification tasks, the computer program
is trained to learn the appropriate function $h:\mathbb{R}^{m}\rightarrow\left\{ 1,2,\cdots,t\right\} .$
Given an input, the learned program determines which of the $t$ categories
the input belongs to via $h$. Deciding if a given picture depicts
a cat or a dog is a canonical example of a classification task.
\item Regression: In regression tasks, the computer program is trained to
predict a numerical value for a given input. The aim is to learn
the appropriate function $h:\mathbb{R}^{m}\rightarrow\mathbb{R}.$
A typical example of regression is predicting the price of a house
given its size, location and other relevant features.
\item Anomaly detection: In anomaly detection (also known as outlier detection)
tasks, the goal is to identify rare items, events or objects that
are significantly different from the majority of the data. A representative
example of anomaly detection is credit card fraud detection where
the credit card company can detect misuse of the customer's card by
modelling his/her purchasing habits.
\item Denoising: Given a noisy example $\tilde{x}\in\mathbb{R}^{n}$, the
goal of denoising is to predict the conditional probability distribution
$P\left(x\vert\tilde{x}\right)$ over noise-free data $x\in\mathbb{R}^{n}$.
\end{itemize}
The measure of the success $P$ of a machine learning algorithm  depends on the task $T$. For example, in the case of classification, $P$ can be measured via the accuracy of the model, i.e. fraction of examples for which the model produces the correct output.
An equivalent description can be in terms of error rate, i.e. fraction
of examples for which the model produces the incorrect output. The goal
of machine learning algorithms is to work well on previously unseen data.
To get an estimate of model performance $P$, it is customary to estimate $P$ on a separate dataset called test set which the machine has not seen during the training.
The data employed for training is known as the training set.

Depending on the kind of experience, $E$, the machine is permitted to
have during the learning process, the machine learning algorithms
can be categorized into supervised learning, unsupervised learning,
and reinforcement learning.
\begin{enumerate}
\item Supervised learning: The goal is to learn a function $y=f(x)$, that returns
the label $y$ given the corresponding unlabeled data $x$. A prominent example
would be images of cats and dogs, with the goal to recognize the
correct animal. The machine is trained with labeled example data,
such that it learns to correctly identify datasets it has not seen
before. Given a finite set of training samples from the join distribution $P(Y, X)$, the task of supervised learning is to infer the probability of a specific label $y$ given example data $x$ i.e., $P(Y=y\vert X=x)$. The function that assigns labels can be inferred from the aforementioned conditional probability distribution.

\item Unsupervised learning: For this type of machine learning, data $x$ is
given without any label. The goal is to recognize possible underlying structures
in the data. The task of the unsupervised machine learning algorithms
is to learn the probability distribution $P\left(x\right)$ or some
interesting properties of the distribution, when given access to several
examples $x$. It is worth stating that the
distinction between supervised and unsupervised learning can be blurry. For example, given a vector $x\in\mathbb{R}^{m},$ the
joint probability distribution can be factorized (using the chain rule
of probability) as
\begin{equation}
P\left(x\right)=\prod_{i=1}^{m}P\left(x_{i}\vert x_{1},\cdots,x_{i-1}\right).\label{eq:factor_chain}
\end{equation}
The above factorization (\ref{eq:factor_chain}) enables us to transform
a unsupervised learning task of learning $P\left(x\right)$ into $m$
supervised learning tasks. Furthermore, given the supervised learning
problem of learning the conditional distribution $P\left(y\vert x\right),$
one can convert it into the unsupervised learning problem of learning
the joint distribution $P\left(x,y\right)$ and then infer $P\left(y\vert x\right)$
via
\begin{equation}
P\left(y\vert x\right)=\frac{P\left(x,y\right)}{\sum_{y1}P\left(x,y1\right)}.\label{eq:infer}
\end{equation}
This last argument suggests that supervised and unsupervised learning are
not entirely distinct. Yet, this distinction between supervised versus unsupervised is sometimes useful for the classification of algorithms.

\item Reinforcement learning: Here, neither data nor labels are available.
The machine has to generate the data itself and improve this data generation process through optimization of a given reward function.  This method is somewhat akin to a human child playing games: The child interacts
with the environment, and initially performs random actions. By external
reinforcement (e.g. praise or scolding by parents), the child learns
to improve itself. Reinforcement learning has shown immense success
recently.  Through reinforcement learning, machines have mastered games that were initially thought to be too complicated for computers to master. This was for example demonstrated by Deepind's AlphaZero which has defeated the best human player in the board game Go \cite{silver2017mastering}.
\end{enumerate}

\begin{figure}[h]
	\includegraphics[width=0.4\textwidth]{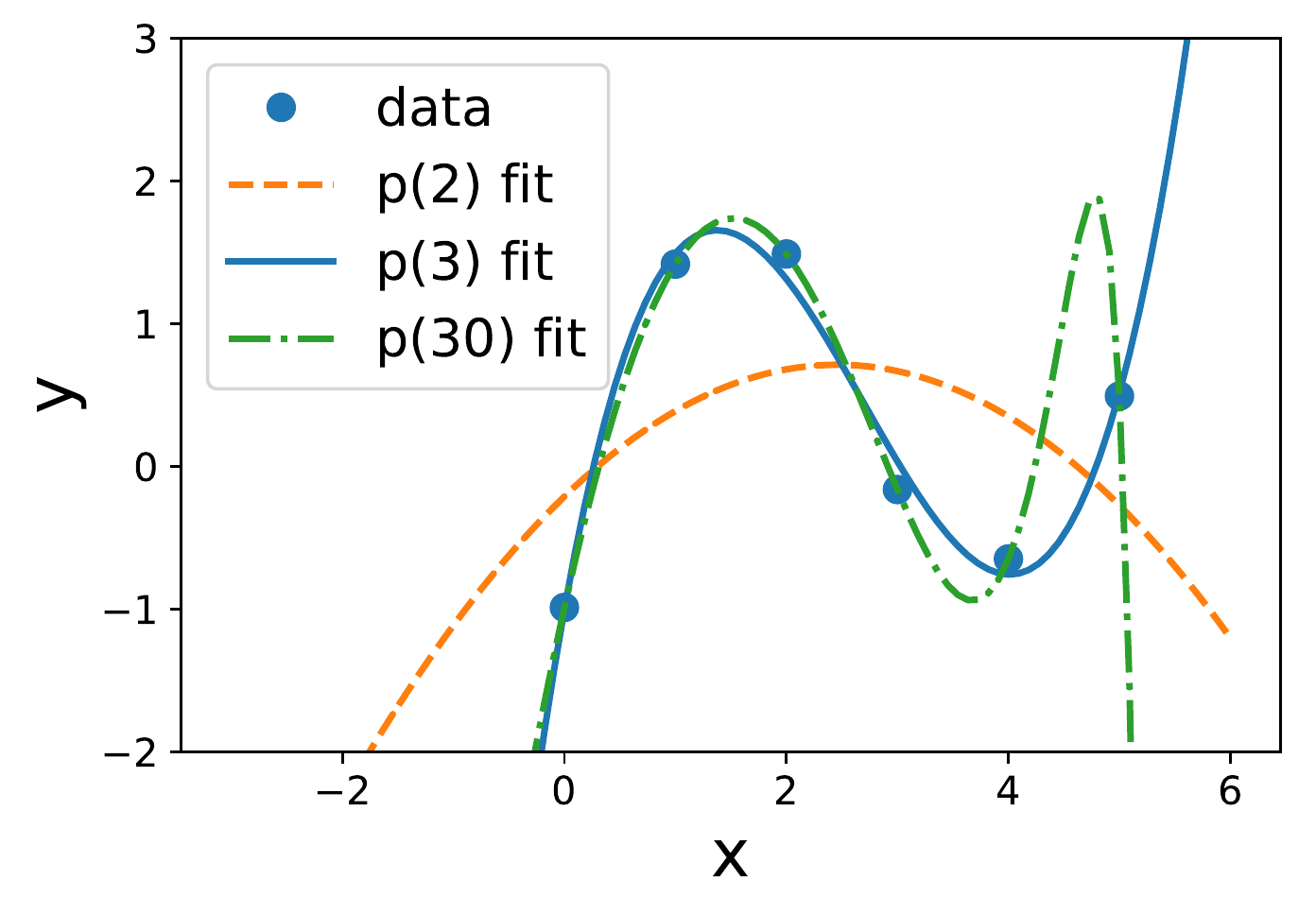}
	\caption{Fitting data sampled from a degree 3 polynomial with small noise. A degree 2 polynomial (orange dashed line) is under-fitting as it has too low model capacity to capture the data. The degree 3 model has the right model capacity (blue solid line). The degree 30 polynomial (red dashed-dotted line) is over-fitting the data, as it captures the sampled data (low training error), but fails to capture the underlying model (high test error).}
	\label{fit}
\end{figure}

One of the central challenges in machine learning is to devise algorithms
that perform well on previously unseen data. The learning ability of a machine to achieve a high performance $P$ on previously unseen data
is called generalization.  The input data to the machine is a set of variables, called features. A specific instance of data is called feature vector.
The error measure on the feature vectors used for training is called training error.  In contrast, the error measure on the test dataset, which the machine has not seen during the training, is
called generalization error or test error. Given an estimate of training
error, how can we estimate test error? The field of statistical learning theory aptly answers this question. The training
and test data are generated according to a probability distribution
over some datasets. Such a distribution is called data-generating distribution.
It is conventional to make independent and identically distributed (IID) assumption, i.e. each example of the dataset is independent of another and, the training and test set are identically distributed. Under such assumptions, the performance of the machine
learning algorithm depends on its ability to reduce the training error
as well as the gap between training and test error. 
If a machine-learning algorithm fails to get sufficiently low training error; the phenomenon
is called under-fitting. On the other hand, if the training error is low, but the test error is large, the phenomenon is called over-fitting. The
capacity of a machine learning model to fit a wide variety of functions
is called model capacity. A machine learning
algorithm with low model capacity is likely to underfit, whereas a too high model capacity often leads to over-fitting the data.
One of the ways to alter the model capacity of a machine learning
algorithm is by constraining the class of functions that the algorithm
is allowed to choose.  For example, to fit a curve to a dataset, one often chooses a set of polynomials as fitting functions (see Fig.\ref{fit}). If the degree of the polynomials is too low, the fit may not be able to reproduce the data sufficiently (under-fitting, orange curve). However, if the degree of the polynomials is too high, the fit will reproduce the training dataset too well, such that noise and the finite sample size is captured in the model (over-fitting, green curve).

\subsection{Artificial Neural Networks}

Most of the recent advances in machine learning were facilitated by
using artificial neurons. The basic structure is a single artificial neuron (AN), which is a real-valued
function of the form, 
\begin{multline}
AN(\boldsymbol{x})=\phi\left(\sum_{i}w_{i}x_{i}\right)\text{ where }\boldsymbol{x}=\left(x_{i}\right)_{i}\in\mathbb{R}^{k},\\
\text{\ensuremath{\boldsymbol{w}=\left(w_{i}\right)_{i}\in\mathbb{R}^{k}} and }\phi:\mathbb{R}\rightarrow\mathbb{R}.\label{eq:AN}
\end{multline}

In equation \ref{eq:AN}, the function $\phi$ is usually known as
activation function. Some of the well-known activation functions are 
\begin{enumerate}
\item Threshold function: $\phi\left(a\right)=1$ if $a>0$ and $0$ otherwise, 
\item Sigmoid function: $\phi\left(a\right)=\sigma(a)=\frac{1}{1+\exp\left(-a\right)}\, ,$
and 
\item Rectified Linear (ReLu) function: $\phi(a)=$ max$\left(0,a\right).$ 
\end{enumerate}
The $\boldsymbol{w}$ vector in equation \ref{eq:AN} is known as
weight vector. 

\begin{figure}[h]
	\includegraphics[width=0.4\textwidth]{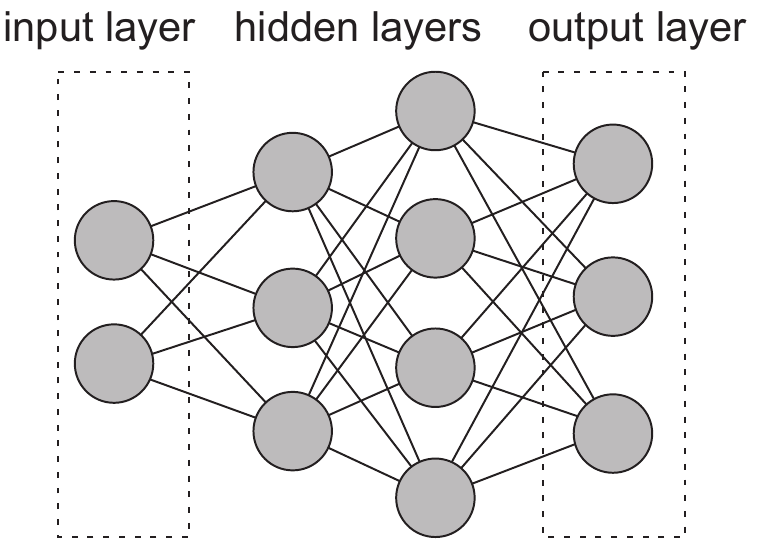}
	\caption{A feed-forward neural networks are a network of artificial neurons chained together in sequence. Each dot corresponds to a artificial neuron, whereas the line indicate the input weights that feed into that neuron. Data fed into the input layer is propagated each layer at a time via the hidden layers (here two hidden layers as example) to the final output layer.}
	\label{FigNN}
\end{figure}

Many of those artificial neurons can be combined together via communication links to perform complex computation. This can be achieved by feeding the output of neurons (weighted by the weights $w_i$) as an input to another neuron, where the activation function is applied again. Such a graph structure $G=\left(V,E\right),$ with
the set of nodes $(V)$ as artificial neurons and edges in $E$ as
connections, is known as an artificial neural network or neural network
in short. The first layer, where the data is fed in, is called input layer. The layers of neurons in-between are the hidden layers, which are defining feature of deep learning. The last layer is called output layer.  A neural network with more than one hidden layer is called deep neural network. Machine learning involving deep neural networks is called Deep learning. A feedforward neural network is a directed acyclic graph, which means that the output of the neurons is fed only into forward direction (see Fig.\ref{FigNN}). 
Apart from the feedforward neural network, some of the popular neural
network architectures are convolutional neural networks (CNN), recurrent
neural networks (RNN), generative adversarial network (GAN), Boltzmann
machine and restricted Boltzmann machines (RBM). We provide a brief
summary of RBM here. For a detailed understanding of various other
neural networks, refer to \cite{goodfellow2016deep}.

\begin{figure}[h]
	\includegraphics[width=0.4\textwidth]{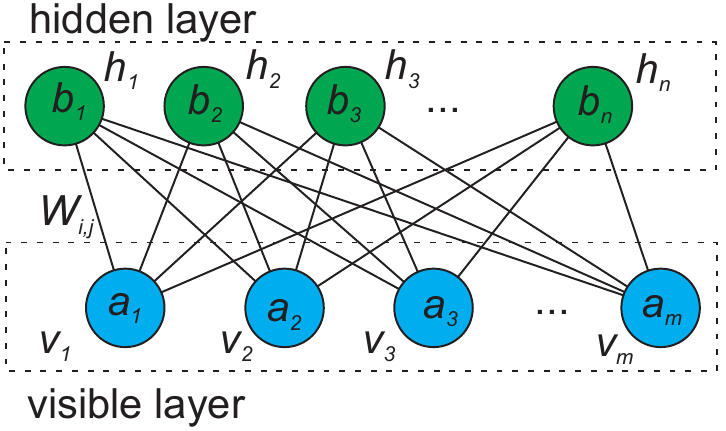}
	\caption{A restricted Boltzman machines (RBM) is composed of hidden and visible layer. Each node has a bias ($a_i$ for visible, $b_j$ for hidden layer), and is connected with weight $W_{i,j}$ to all the nodes in the opposite layer. There are no intra-layer connections.}
	\label{FigRBM}
\end{figure}

An RBM is a bipartite graph with two kinds of binary-valued neuron
units, namely visible $\left(\boldsymbol{v}=\left\{ v_{1},v_{2},\cdots\right\} \right)$
and hidden $\left(\boldsymbol{h}=\left\{ h_{1},h_{2},\cdots\right\} \right)$ (see Fig.\ref{FigRBM})
The weight matrix $W=\left(w_{ij}\right)$ encodes the weight corresponding
to the connection between visible unit $v_{i}$ and hidden unit $h_{j}$~\cite{hinton2012practical}.
Let the bias weight (offset weight) for the visible unit $v_{i}$
be $a_{i}$ and hidden unit $h_{j}$ be $b_{j}$. 

For a given configuration
of visible and hidden neurons $\left(\boldsymbol{v,h}\right)$, one
can define an energy function $E\left(\boldsymbol{v,h}\right)$ inspired from statistical models for spin systems as
\begin{equation}
E\left(\boldsymbol{v,h}\right)=-\sum_{i}a_{i}v_{i}-\sum_{j}b_{j}h_{j}-\sum_{i}\sum_{j}v_{i}w_{ij}h_{j}.\label{eq:RBM_energy}
\end{equation}
The probability of a configuration $\left(\boldsymbol{v,h}\right)$
is given by Boltzmann distribution,
\begin{equation}
P\left(\boldsymbol{v,h}\right)=\frac{\exp\left(-E\left(\boldsymbol{v,h}\right)\right)}{Z},\label{eq:prob_config}
\end{equation}
where $Z=\sum_{\boldsymbol{v,h}}\exp\left(-E\left(\boldsymbol{v,h}\right)\right)$
is a normalization factor, commonly known as partition function. 
As there are no intra-layer connections, one can sample from this distribution easily. Given a set of visible units $\boldsymbol{v}$, the probability of a specific hidden unit being $h_j=1$ is 
\begin{equation}
p(h_j=1| \boldsymbol{v})=\sigma(h_j + \sum_i w_{i,j}v_i)\, ,
\end{equation}
where $\sigma(a)$ is the sigmoid activation function as introduced earlier. A similar relation holds for the reverse direction, e.g. given a set of hidden units, what is the probability of the visible unit.

With these concepts, complicated probability distributions $P\left(\boldsymbol{v}\right)$ over some variable  $\boldsymbol{v}$ can be encoded and trained within the RBM.
Given a set of training vectors $\boldsymbol{v}\in\mathbb{V}$, the
goal is find the weights $W^{\prime}$ of the RBM that fit the training set best
\begin{equation}
\text{arg max}_{W^{\prime}}\prod_{\boldsymbol{v}\in\mathbb{V}}P\left(\boldsymbol{v}\right).\label{eq:training_RBM}
\end{equation}
RBMs have been shown to be a good Ansatz to represent many-body wavefunctions, which are difficult to handle with other methods due to the exponential scaling of the dimension of the Hilbert space. This method has been successfully applied to quantum many-body physics
and quantum foundations problems \cite{deng2018machine,carleo2017solving}.

In context beyond physics, machine learning with deep neural networks has accomplished many significant
milestones, such as mastering various games, image recognition, self-driving
cars, and so forth. Its impact on the physical sciences is just about
to start \cite{carleo2019machine,deng2018machine}.

\subsection{Relation with Artificial Intelligence}

The term ``artificial intelligence'' was first coined in the famous
1956 Dartmouth conference \cite{mccarthy2006proposal}. Though the term was invented in 1956, the
operational idea can be traced back to Alan Turing's influential ``Computing Machinery and Intelligence''
paper in 1950, where Turing asks if a machine can think \cite{turing2009computing}. The idea of designing machines that can think dates back to ancient Greece. Mythical characters such as Pygmalion, Hephaestus and Daedalus can be interpreted as some of the legendary inventors and Galatea, Pandora and Hephaestus can be thought of as examples of artificial life \cite{goodfellow2016deep}. The field
of artificial intelligence is difficult to define, as can be seen
by four different and popular candidate definitions \cite{russell2002artificial}. The definitions
start with ``AI is the discipline that aims at building ...''
\begin{enumerate}
\item (Reasoning-based and human-based): agents that can reason like humans.
\item (Reasoning-based and ideal rationality): agents that think rationally.
\item (behavior-based and human-based): agents that behave like humans.
\item (behavior-based and ideal rationality): agents that behave rationally.
\end{enumerate}
Apart from its foundational impact in attempting to understand “intelligence,” AI has reaped practical impacts such as automated routine labour and automated medical diagnosis, to name a few among many. The real challenge of AI is to execute tasks which are easy for people to perform, but hard to express formally. An approach to solve this problem is by allowing machines to learn from experience, i.e. via machine learning. From a foundational point of judgment, the study of machine learning is vital as it helps us understand the meaning of “intelligence”. It is worthwhile mentioning that deep learning is a subset of machine learning which can be thought of as a subset of AI (see Fig.\ref{Fig0}).
\begin{figure}[t]
	\includegraphics[width=0.25\textwidth]{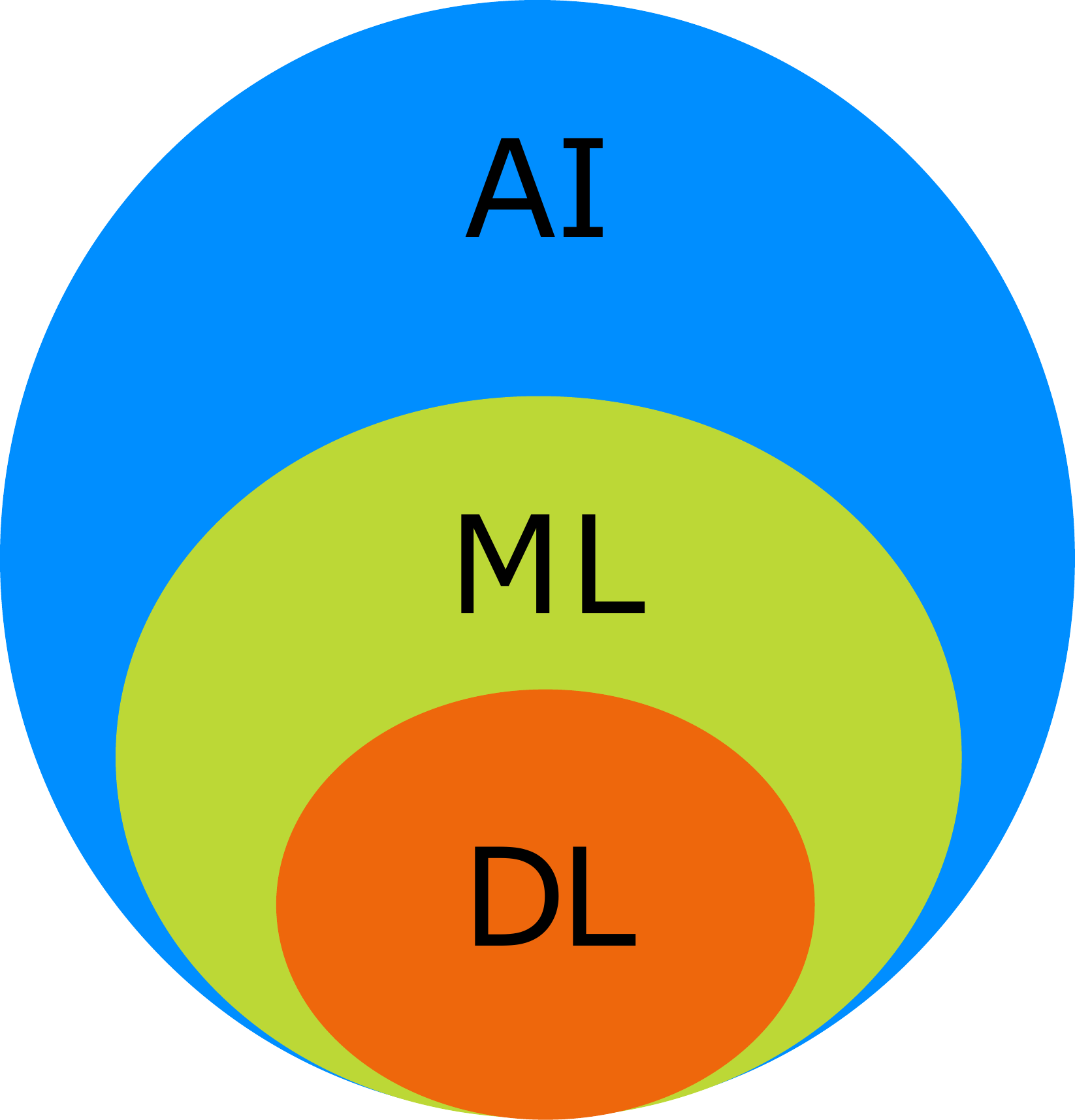}
	\caption{Deep learning (DL) is a sub-discipline of machine learning (ML). ML can be considered a sub-discipline of artificial intelligence (AI).}
	\label{Fig0}
\end{figure}
\section{\label{sec:Foundation} Quantum Foundations}

The mathematical edifice of quantum theory has intrigued and puzzled physicists, as well as philosophers, for many years. Quantum foundations seeks to understand and develop the mathematical as well as conceptual understanding of quantum theory. The study concerns the search for non-classical
effects such as Bell nonlocality, contextuality and different interpretations of
quantum theory.  This study also involves the investigation of physical principles that can put the theory into an axiomatic framework, together with an exploration of possible extensions of quantum theory. In this survey, we will
focus on non-classical features such as Entanglement, Bell nonlocality, contextuality
and quantum steering in some detail. 

An interpretation of quantum theory can be viewed as a map from the elements of the mathematical structure of quantum theory to elements of reality. Most of the interpretations of quantum theory
seek to explain the famous measurement problem. Some of the brilliant interpretations are the Copenhagen interpretation, quantum Bayesianism \cite{fuchs2011quantum},
the many-world formalism \cite{everett1957relative,dewitt2015many} and the consistent history interpretation \cite{omnes1992consistent}. The axiomatic reconstruction of quantum theory is generally categorized into two parts: the generalized
probabilistic theory (GPT) approach \cite{barrett2007information} and the black-box approach \cite{acin2017black}. The
physical principles underlying the framework for axiomatizing
quantum theory are non-trivial communication complexity \cite{van2013implausible, brassard2006limit}, information
causality \cite{pawlowski2009information}, macroscopic locality~\cite{navascues2010glance}, negating the advantage for nonlocal computation \cite{linden2007quantum},
consistent exclusivity \cite{yan2013quantum} and local orthogonality \cite{fritz2013local}. The extensions of
the quantum theory include collapse models \cite{mcqueen2015four}, quantum measure theory \cite{sorkin1994quantum} and acausal quantum processes \cite{oreshkov2012quantum}.
\subsection{\label{subsec:Ent} Entanglement}

Quantum interaction inevitably leads to quantum entanglement. The individual states of two classical systems after an interaction are independent of each other. Yet, this is not the case for two quantum systems \cite{raimond2001manipulating, horodecki2009quantum}.  Quantum states can be pure or mixed. For a bipartite pure quantum state $|\psi\rangle \in \cal{H}^A \otimes \cal{H}^B$ is entangled if it cannot be written as a product state, i.e. $|\psi \rangle = |\psi^A\rangle \otimes |\psi^B\rangle$ for some $|\psi^A \rangle \in \cal{H}^A$ and $|\psi^B\rangle \in \cal{H}^B$. A mixed bipartite state is however expressed in terms of a density matrix, $\rho$. Like for pure state, a density matrix $\rho$ is entangled if it cannot be expressed in the form $\displaystyle \rho = \sum_i p_i |\psi_i^A \rangle$ $\langle \psi_i^A| \otimes  |\psi_i^ \rangle$ $\langle \psi_i^B|$. An $n-$partite state
$\rho_{sep}$ is called separable if it can be represented as a convex
combination of product states i.e. 
\begin{equation}
\rho_{sep}=\sum_{i}p_{i}\rho_{i}^{1}\otimes\rho_{i}^{2}\otimes\cdots\otimes\rho_{i}^{n},\label{eq:sep}
\end{equation}
where $0\leq p_{i}\leq1$ and $\sum_{i}p_{i}=1.$
A quantum state is called separable if it is not entangled.

Computationally it is not easy to check if a mixed state, especially in higher dimensions and for more parties, is separable or entangled.   Numerous measures of quantum entanglement for both pure and mixed states are proposed \cite{horodecki2009quantum}: for bipartite systems where the dimension of $\mathcal{H}^i (i = A,B)$ is 2, a good measure is concurrence \cite{wootters2001entanglement}.  Other measures for quantifying entanglement are  entanglement of formation, robustness of entanglement, Schmidt rank, squashed entanglement and so forth \cite{horodecki2009quantum}. It turns out that for any entangled state $\rho$, there exists a Hermitian matrix $W$ such that $\text{Tr} (\rho W) <0 $ and for all separable states $\rho_\text{sep}$,  $\text{Tr} (\rho W) \geq 0 $ ~\cite{horodecki1996necessary,terhal2002detecting,krammer2009multipartite,ekert2002direct}. 

Quantum entanglement was mooted a long time ago by Erwin Schr\"odinger, but it took another thirty years or so for John Bell to show that quantum theory imposes strong constraints on statistical correlations in experiments. Yet, correlations are not tantamount to causation and one wonders if machine learning could do better \cite{pearle2000causality}. In the sixties and seventies, this Gedanken experiment was given further impetus with some ingenious experimental designs \cite{freedman1972experimental,kocher1967polarization,aspect1981experiences,aspect1982experimental}.  The advent of quantum information in the nineties gave a further push: quantum entanglement became a useful resource for many quantum applications, ranging from teleportation \cite{bennett1993teleporting}, communication \cite{ekert1991quantum}, purification \cite{bennett1996purification}, dense coding \cite{bennett1992communication} and computation \cite{deutsch1985quantum,feynman2018feynman}. Interestingly, quantum entanglement is a fascinating area that emerges almost serendipitously from the foundation of quantum mechanics into real practical applications in the laboratories.

\subsection{\label{subsec:Bell} Bell Nonlocality}

\begin{figure}[t]
	\includegraphics[width=0.35\textwidth]{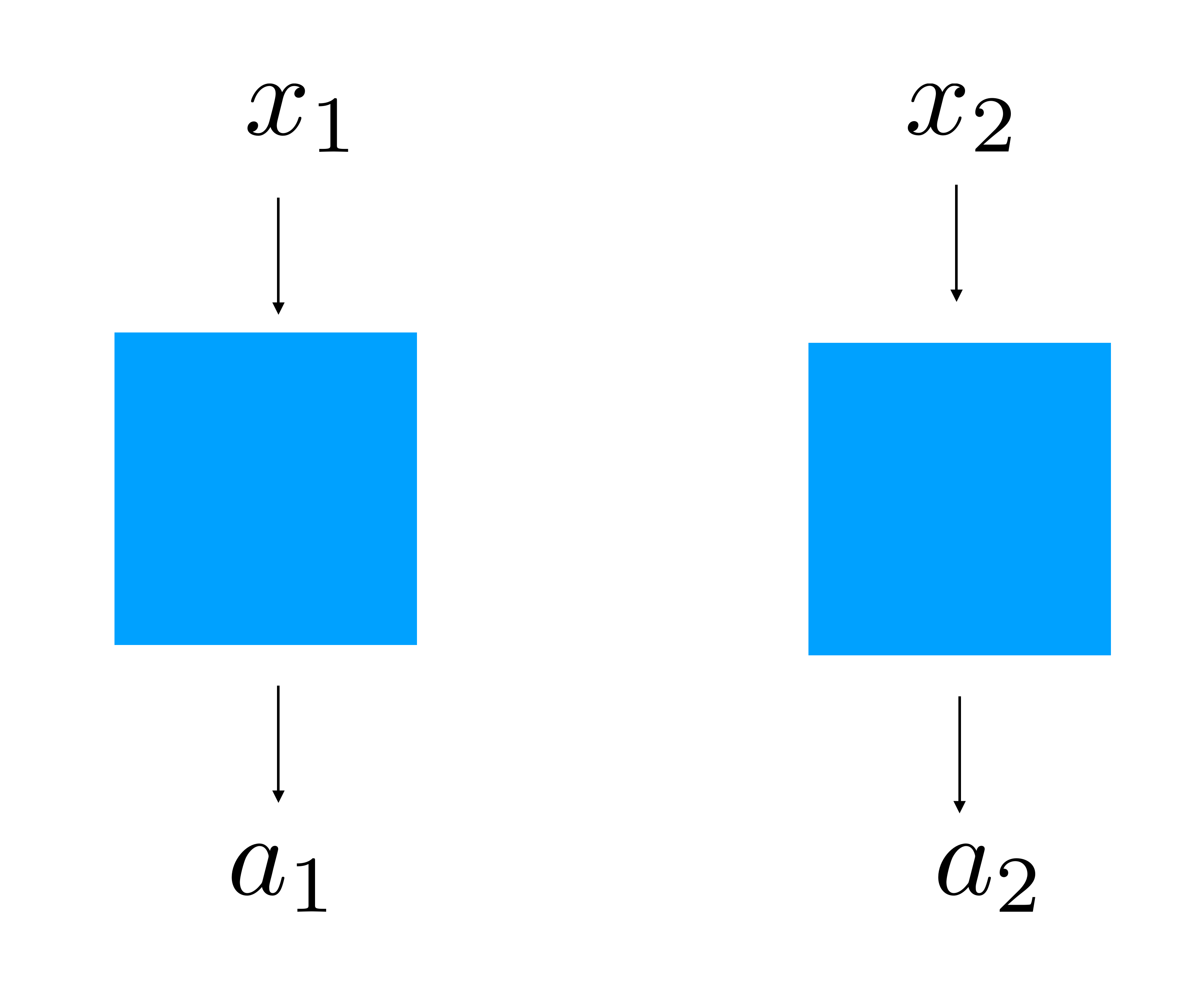}
	\caption{The simplest scenario in which Bell nonlocality can be demonstrated is the famous Clauser-Horn-Shimony-Holt
(CHSH) experiment. There are two parties involved. Each party can perform two dichotomic measurements. Thus, it is a $(2,2,2)$ scenario. The experimental statistics corresponds to probabilities of the form $P\left(a_1,a_2 \vert x_1, x_2\right).$ }
	\label{CHSH_Scenario}
\end{figure}

According to John Bell \cite{Bell64}, any theory based on the collective
premises of locality and realism must be at variance with experiments
conducted by spatially separated parties involving shared entanglement, if the underlying measurement events are space-like separated.
The phenomenon, as discussed before, is known as Bell nonlocality \cite{brunner2014bell}. Apart from its significance in understanding foundations
of quantum theory, Bell nonlocality is a valuable resource for many
emerging device-independent quantum technologies like quantum key
distribution (QKD), distributed computing, randomness certification
and self-testing \cite{ekert1991quantum,pironio2010random,Yao_self}.
The experiments which can potentially manifest Bell nonlocality are
known as Bell experiments. A Bell experiment involves $N$ spatially
separated parties $A_{1},A_{2},\cdots,A_{N}$. Each party receives
an input $x_{1},x_{2},x_{3,}\cdots,x_{N}\in\mathcal{X}$ and gives
an output $a_{1},a_{2},a_{3},\cdots,a_{N}\in\mathcal{A}.$ For various
input-output combinations one gets the statistics of the following
form: 
\begin{equation}
\mathcal{P}=\left\{ P\left(a_{1},a_{2},\cdots,a_{N}\vert x_{1,}x_{2},\cdots,x_{N}\right)\right\} _{x_{1},\cdots,x_{N}\in\mathcal{X},a_{1},\cdots,a_{N}\in\mathcal{A}}.\label{eq:behavior}
\end{equation}
We will refer to $\mathcal{P}$ as behavior. A Bell experiment involving
$N$ space-like separated parties, each party having access to $m$
inputs and each input corresponding to $k$ outputs is referred to
as $\left(N,m,k\right)$ scenario. The famous Clauser-Horn-Shimony-Holt
(CHSH) experiment is a $(2,2,2)$ scenario \cite{CHSH}, and it is the simplest
scenario in which Bell nonlocality can be demonstrated (see Fig. \ref{CHSH_Scenario}). 

A behavior $\mathcal{P}$ admits local hidden variable description if and only if 
\begin{multline}
P\left(a_{1},a_{2},\cdots,a_{N}\vert x_{1,}x_{2},\cdots,x_{N}\right)=\sum_{\lambda}P\left(\lambda\right)P\left(a_{1}\vert x_{1}, \lambda\right) \\\cdots  P\left(a_{n}\vert x_{n},\lambda \right).\label{eq:LHV}
\end{multline}
\begin{figure}[t]
	\includegraphics[width=0.45\textwidth]{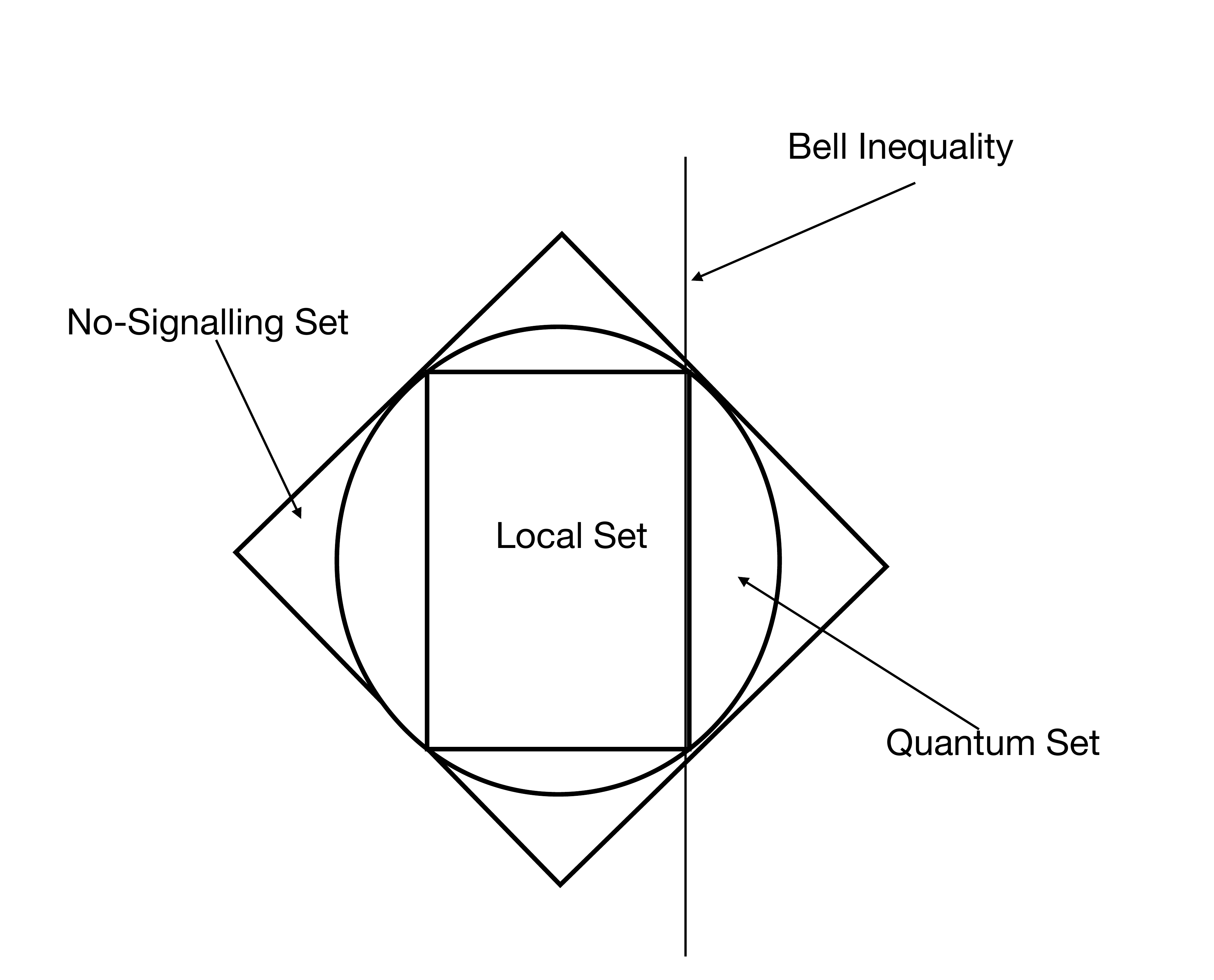}
	\caption{The set of local behaviors ($\mathcal{L}$) forms a convex polytope (bounded polyhedron). The set of quantum behaviors $\mathcal{Q}$ is a convex set. The set of behaviors compatible with no-signalling condition ($\mathcal{NS}$) is again a convex polytope. It is important to note $\mathcal{L} \subseteq \mathcal{Q} \subseteq \mathcal{NS}.$ The hyperplanes defining the boundary of the local set $\mathcal{L}$ are Bell inequalities. It is important to reiterate that the violation of Bell inequalities means that a local realistic description of nature is impossible.}
	\label{Polytope}
\end{figure}
This is known as local behavior. The set of local behaviors  $(\mathcal{L})$
forms a convex polytope, and the facets of this polytope are Bell
inequalities (see Fig. \ref{Polytope}). In quantum theory, Born rule governs probability according
to 
\begin{equation}
P\left(a_{1},a_{2},\cdots,a_{N}\vert x_{1,}x_{2},\cdots,x_{N}\right)=\text{Tr}\left[\left[M_{a_{1}\vert x_{1}}\otimes\cdots\otimes M_{a_{N}\vert x_{N}}\right]\rho\right],\label{eq:Quant}
\end{equation}
where $\left\{ M_{a_{i}\vert x_{i}}\right\} _{i}$ are positive-operator
valued measures (POVMs) and $\rho$ is a shared density matrix. If
a behavior satisfying Eq.\ref{eq:Quant} falls outside $\mathcal{L}$,
it then violates at least one Bell inequality, and such behavior
is said to manifest Bell nonlocality. The condition that parties do not communicate during the course of the Bell experiment is known as no-signalling condition. Intuitively speaking, it means that the choice of input of one party can not be used for signalling among parties. Mathematically it means,
\begin{align}
    \sum_{a_j} P\left(a_i,a_j \vert x_i,x_j\right) = P\left(a_i \vert x_i \right) \forall i,j \in \{1,\cdots,N\} \text{ and }  i\neq j.
\end{align}
The set of behaviours which satisfy the no signalling condition are known as no-signalling behaviours. We will denote the aforementioned set by $\mathcal{NS}$. The no-signalling set also forms a polytope. Furthermore, $\mathcal{L} \subseteq \mathcal{Q} \subseteq \mathcal{NS}$ (see Fig. \ref{Polytope}). The no-signalling behaviours which do not lie in $\mathcal{Q}$ are also known as post-quantum behaviours. In the (2,2,2) scenario i.e. CHSH Scenario, there is a unique Bell inequality, namely CHSH inequality, upto relabelling of inputs and outputs. The CHSH inequality is given by
\begin{equation}
    E_{0,0} +  E_{0,0} +  E_{0,0} +  E_{0,0}  \leq 2
    \label{CHSH_in}
\end{equation}
where $E_{x_1,x_2} = P\left(a_1 = a_2 \vert x_1 = x_2 \right) - P\left(a_1 \neq a_2 \vert x_1 = x_2 \right).$ All local hidden variable theories satisfy CHSH inequality. In quantum theory, suitably chosen measurement settings and state can lead to violation of CHSH inequality and thus the CHSH inequality can be used to witness Bell nonlocal nature of quantum theory. Quantum behaviours achieve upto $2\sqrt{2},$ known as the Tsirelson bound. The upper bound for no-signalling behaviours (no-signalling bound) on the CHSH inequality is $4$.

\subsection{\label{subsec:Context} Contextuality}
An intuitive feature of classical models is non-contextuality which means that any measurement has a value independent of other compatible measurements it is performed together with. A set of compatible measurements is called context. It was shown by Simon Kochen and Ernst Specker (as well as John Bell) \cite{kochen1967problem} that non-contextuality conflicts with quantum theory.

\begin{figure}[t]
	\includegraphics[width=0.45\textwidth]{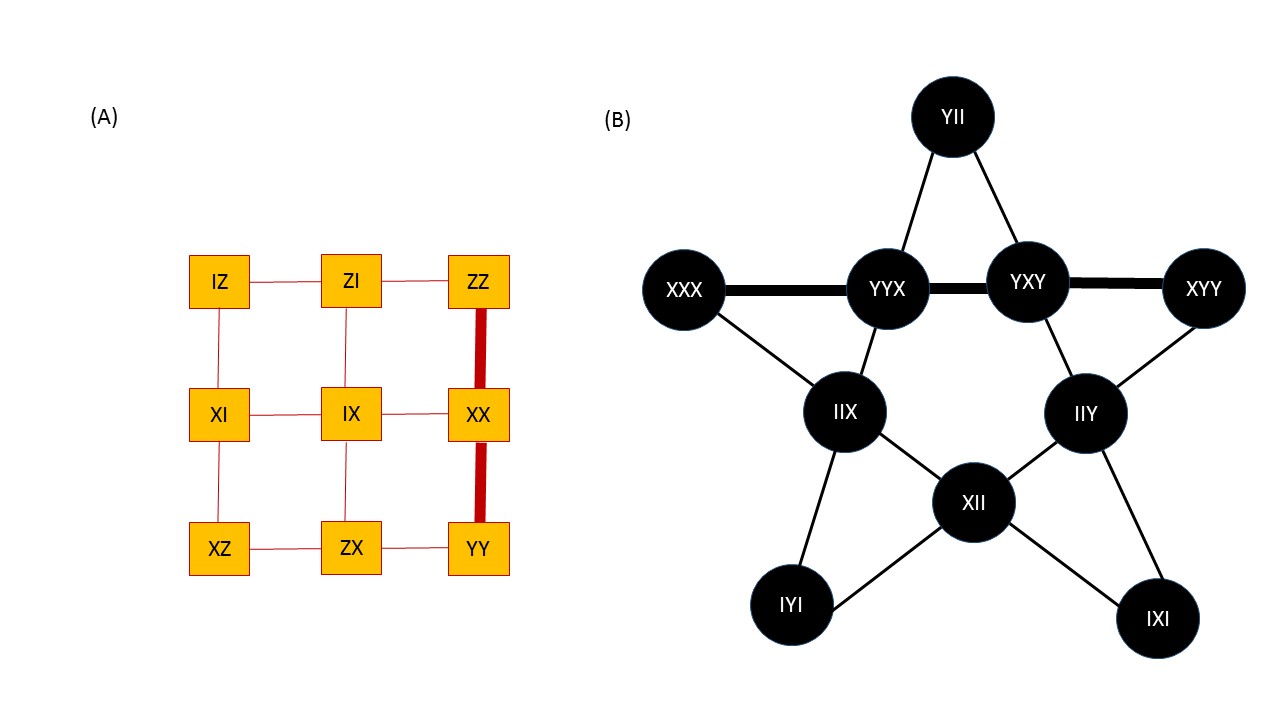}
	\caption{The Mermin-Peres square in (A) provides a proof of Kochen-Specker theorem.  Each line in the $3 \times 3$ grid describes mutually commuting operators.  Since the square of the operator at each node is the identity matrix, its eigenvalues are $\pm 1$. The product of the operator along each line is the Identity, except for the bold one in which the product is minus the Identity matrix.There is no way to assign definite value $\pm 1$ to each operator to get these product rules since each operator appears in exactly two lines (or contexts). (B) shows the Mermin's pentagram where we have four mutually commuting three-qubit operators. Along each line the product of the operators is Identity except for the bold one, leading to a contradiction. Classically, it is impossible to assign the numbers $\pm$ to each node such that the product of the numbers along a straightline is 1, except for the bold line which is -1. }
	\label{fig:merminsq}
\end{figure}

The contextual nature of quantum theory can be established via simple constructive proofs.  In Fig. \ref{fig:merminsq} (A), one considers an array of operators on a two-qubit system in any quantum state.  There are nine operators, and each of them has eigenvalue $\pm 1$.  The three operators in each row or column commute and so it is easy to check that each operator is the product of the other two operators on a particular row or column, with a single exception, the third operator in the third column equals the minus of the product of the other two. Suppose there exist pre-assigned values ($-1$ or $+ 1$) for the outcomes of the nine operators, then we can replace the none operators by the pre-assigned values. However, there is no consistent way to assign such values to the nine operators so that the product of the numbers in every row or column (except for the operators along the bold line) yield 1 (the product yields -1). Notice that each operator (node) appears in exactly two lines or context. Kochen and Specker provided the first proof of quantum contextuality with a complicated construct involving 117 operators on a 3-dimensional space \cite{kochen1967problem}. Another example is the pentagram \cite{mermin1993hidden} in Fig. \ref{fig:merminsq}.

Bell nonlocality is regarded as a particular case
of contextuality where the space-like separation of parties involved
creates ''context'' \cite{CSW,amaral2018graph}. The study of contextuality has led not only to insights into the foundations of quantum mechanics, but it also offers practical implications as well \cite{Cabello_QKD, bharti2019non,raussendorf2013contextuality,bharti2019robust,mansfield2018quantum,saha2019state,bharti2019local, delfosse2015wigner,singh2017quantum,pashayan2015estimating,bharti2018simple,bermejo2017contextuality,catani2018state,arora2019revisiting,Howard2014}.
There are several frameworks for
contextuality including sheaf-theoretic framework \cite{abramsky2011sheaf}, graph and hypergraph
framework~\cite{CSW,acin2015combinatorial}, contextuality-by-default framework \cite{dzhafarov2017contextuality, dzhafarov2019joint, kujala2019measures} and operational framework \cite{spekkens2005contextuality}.
A scenario exhibits contextuality if it does not admit the non-contextual
hidden variable (NCHV) model \cite{KS67}.

\subsection{\label{subsec:Steer} Quantum Steering}

Correlations produced by steering lie between the Bell nonlocal
correlations and those generated from entangled states \cite{quintino2015inequivalence,wiseman2007steering}. A state that
manifests Bell nonlocality for some suitably chosen measurement settings
also exhibits steering \cite{schrodinger1935discussion}. Furthermore, a state which exhibits steering
must be entangled. A state demonstrates steering if it does not admit ``local
hidden state (LHS)'' model \cite{wiseman2007steering}. We discuss this formally here.

Alice and Bob share some unknown quantum state $\rho^{AB}.$ Alice
can perform a set of POVM measurements $\left\{ M_{a}^{x}\right\} _{a}.$
The probability of her getting outcome $a$ after choosing measurement
$x$ is given by 
\begin{multline}
P\left(a\vert x\right)=\text{Tr}\left[\left(M_{a}^{x}\otimes I\right)\rho^{AB}\right]=\text{Tr}\left\{ \text{Tr}_{A}\left[\left(M_{a}^{x}\otimes I\right)\rho^{AB}\right]\right\} \\ 
=\text{Tr}\left[\rho_{a\vert x}^{B}\right],\label{eq:Steer_meas}
\end{multline}
where $\rho_{a\vert x}^{B}=\text{Tr}_{A}\left[\left(M_{a}^{x}\otimes I\right)\rho^{AB}\right]$
is Bob's residual state upon normalization. A set of operators $\left\{ \rho_{a\vert x}^{B}\right\} _{a,x}$
acting on Bob's space is called an assemblage if 
\begin{equation}
\sum_{a}\rho_{a\vert x}^{B}=\sum_{a}\rho_{a\vert x^{\prime}}^{B}\quad\forall x\neq x^{\prime}\label{eq:Steer_NS}
\end{equation}
and
\begin{equation}
\sum_{a}\text{Tr}\left[\rho_{a\vert x}^{B}\right]=1\quad\forall x.\label{eq:Steer_normal}
\end{equation}
Condition \ref{eq:Steer_NS} is the analogue of the no-signalling condition.
An assemblage $\left\{ \rho_{a\vert x}^{B}\right\} _{a,x}$ is said
to admit LHS model if there exists some hidden variable $\lambda$
and some quantum state $\rho_{\lambda}$ acting on Bob's space such
that
\begin{equation}
\rho_{a\vert x}^{B}=\sum_{\lambda} P\left(\lambda\right)P\left(a\vert x,\lambda\right)\rho_{\lambda}.\label{eq:LHS}
\end{equation}
A bipartite state $\rho^{AB}$ is said to be steerable from Alice
to Bob if there exist measurements for Alice such that the corresponding
assemblage does not satisfy equation \ref{eq:LHS}. Determining whether
an assemblage admits LHS model is a semidefinite program (SDP). The concept of steering is asymmetric by definition, i.e. even if Alice could steer Bob's state, Bob may not be able to steer Alice's state.

\section{\label{sec:Oracle} Neural network as Oracle for Bell Nonlocality}

The characterization of the local set for the convex scenario via
Bell inequalities becomes intractable as the complexity of the underlying
scenario grows (in terms of the number of parties, measurement settings and outcomes). For networks where several independent sources are shared
among many parties, the situation gets increasingly worse. The local set
is remarkably non-convex, and hence proper analytical and numerical
characterization, in general, is lacking. Applying machine learning technique to tackle these issues were studied by Canabarro {\em et al.}~\cite{canabarro2019machine} and
Krivachy {\em et al.} \cite{krivachy2019neural}.
In the work by Canabarro {\em et al.}, the detection and characterization of nonlocality is done through an ensemble of multilayer perceptrons
blended with genetic algorithms (see \ref{sec:MLandCorr}). 

\subsection{\label{sec:MLandCorr}Machine Learning Nonlocal Correlations}

Given a behavior, deciding whether it is classical or non-classical
is an extremely challenging task since the underlying scenario grows in complexity very quickly.
Canabarro \textit{et al.}~\cite{canabarro2019machine} use supervised machine learning with an ensemble
of neural networks to tackle the approximate version of the problem
(i.e. with a small margin of error) via regression. The authors ask
``How far is a given correlation from the local set.'' The input feature vector to the neural network is a random correlation vector. For a given behavior, the output
(label) is the distance of the feature vector from the classical, i.e. local set. The nonlocality quantifier of a behavior $q$ is the minimum
trace distance, denoted by $NL(q)$ \cite{brito2018quantifying}. For the two-party scenario, the
nonlocality quantifier is given by
\begin{equation}
NL(q)\equiv\frac{1}{2\vert x\vert\vert y\vert}\text{min}_{p\in\mathcal{L}}\sum_{a,b,x,y}\vert q-p\vert,\label{eq:NLQ}
\end{equation}
where $\mathcal{L}$ is the local set and $\vert x\vert=\vert y\vert=m$
is the input size for the parties. The training points are generated
by  sampling the set of non-signalling behaviors randomly and then
calculating its distance from the local set via equation \ref{eq:NLQ}.
Given a behavior $q$, the distance predicted by the neural network
is never equal to the actual distance, i.e. there is always a small
error $(\epsilon\neq$0). Let us represent the learned hypothesis
as $f:q\rightarrow\mathbb{R}.$ The performance metric of the learned
hypothesis $f$ is given by
\begin{equation}
\mathcal{P}(f)\equiv\frac{1}{N}\sum_{i=1}^{N}\vert NL\left(q_{i}\right)-f\left(q_{i}\right)\vert.\label{eq:P(M)}
\end{equation}
In experiments such as entanglement swapping,
which comprises three separated parties sharing two independent sources
of quantum states, the local set admits the following form and the set is
non-convex, 
\begin{multline}
P\left(a_{1},a_{2},a_{3}\right)=\sum_{\lambda_{1},\lambda_{2}}P\left(\lambda_{1}\right)P\left(\lambda_{2}\right)P\left(a_{1}\vert x_{1},\lambda_{1}\right)P\left(a_{2}\vert x_{2},\lambda_{2}\right)\\ P\left(a_{3}\vert x_{3},\lambda_{3}\right).\label{eq:BL}
\end{multline}
Here, non-convexity emerges from the independence of the sources i.e.
$P\left(\lambda_{1},\lambda_{2}\right)=P\left(\lambda_{1}\right)P\left(\lambda_{2}\right).$

\begin{figure}[t]
	\includegraphics[width=0.5\textwidth]{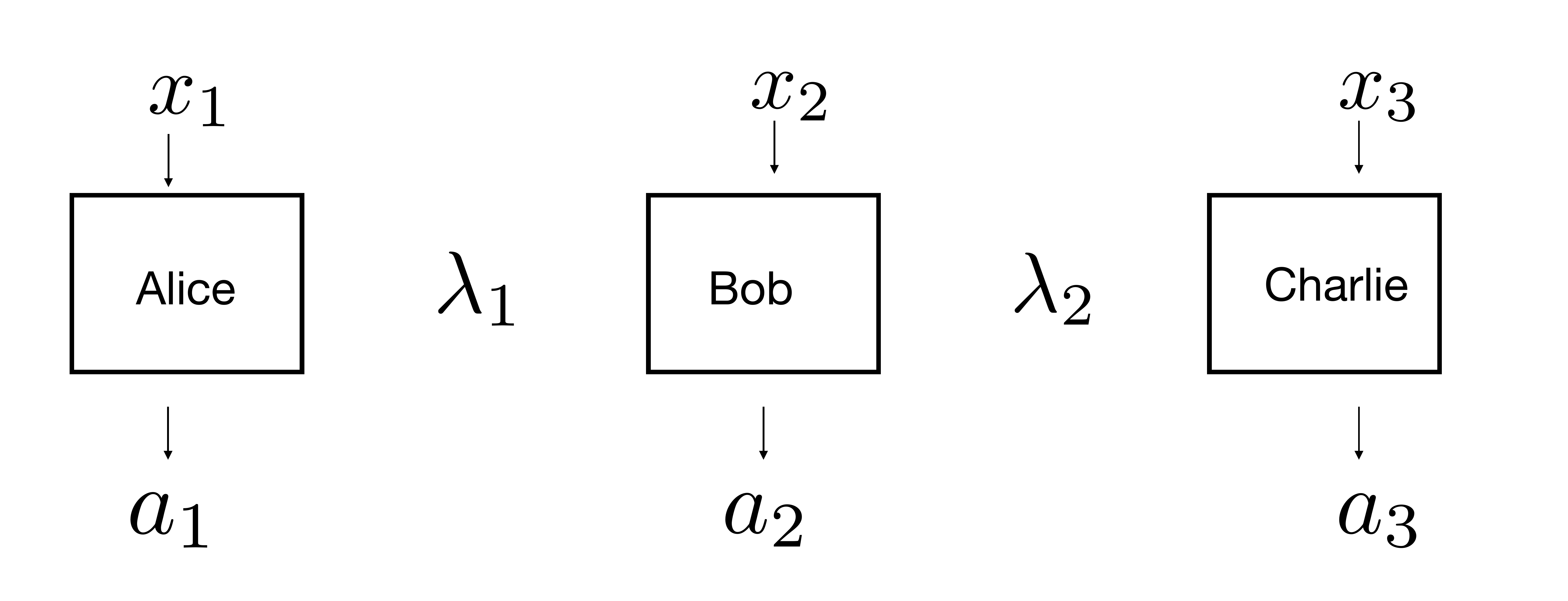}
	\caption{There are three separated parties with two sources of hidden variables. The sources are independent $P\left(\lambda_{1},\lambda_{2}\right)=P\left(\lambda_{1}\right)P\left(\lambda_{2}\right).$ The local set is no longer convex. In experiments such as entanglement swapping, such non-convexity arise. Deciding whether a behaviour is classical or quantum gets complicated
for such scenarios. }
	\label{Bilocal_Scenario}
\end{figure}
In this case, the Bell inequalities are no longer linear. Characterizing
the set of classical as well as quantum behaviors gets complicated
for such scenarios \cite{tavakoli2014nonlocal, branciard2010characterizing}.  

Canabarro \textit{et al.} train the model for convex as well as non-convex
scenarios. They also train the model to learn post-quantum correlations.
The techniques studied in the paper are valuable for understanding Bell nonlocality
for large quantum networks, for example those in quantum internet.

\subsection{\label{sec:MLandNetworks} Oracle for Networks}

Given an observed probability distribution corresponding to scenarios
where several independent sources are distributed over a network,
deciding whether it is classical or non-classical is an important
question, both from practical as well as foundational viewpoint. The
boundary separating the classical and non-classical correlations is
extremely non-convex and thus a rigorous analysis is exceptionally challenging.

In reference \cite{krivachy2019neural}, the authors encode the causal
structure into the topology of a neural network and numerically determine
if the target distribution is ``learnable''. A behavior belongs
to the local set if it is learnable. The authors harness the fact
that the information flow in feedforward neural networks and causal
structures are both determined by a directed acyclic graph. The topology
of the neural network is chosen such that it respects the causality
structure. The local set corresponding to even elementary causal networks
such as triangle network is profoundly non-convex, and thus analytical characterization
of the same is a notoriously tricky task. Using the neural network
as an oracle, Krivachy \textit{et al.} \cite{krivachy2019neural} convert the membership in a local
set problem to a learnability problem. For a neural network with adequate
model capacity, a target distribution can be approximated if it is
local. The authors examine the triangle network with quaternary outcomes
as a proof-of-principle example. In such a scenario, there are three
independent sources, say $\alpha,\beta$ and $\gamma$. Each of the
three parties receives input from two of the three sources and process
the inputs to provide outputs via fixed response functions. The outputs
for Alice, Bob and Charlie will be indicated by $a,b,c\in\left\{ 0,1,2,3\right\} .$
The scenario as discussed here can be characterized by the probability
distribution $P\left(a,b,c\right)$ over the random variables $a,b$
and $c$. If the network is classical, then the distribution can be
represented by a directed acyclic graph known as a Bayesian network
(BN). 

Assuming the distribution $P\left(a,b,c\right)$ over the random
variables $a,b$ and $c$ to be classical, it is assumed  without loss of
generality that the sources send a random variable 
drawn uniformly from the interval $0$ to $1.$ A classical distribution for
such a case admits the following form:
\begin{equation}
P\left(a,b,c\right)=\int_{0}^{1}d\alpha d\beta d\gamma P_\text{A}\left(a\vert\beta,\gamma\right)P_\text{B}\left(b\vert\gamma,\alpha\right)P_\text{C}\left(c\vert\alpha,\beta\right).\label{eq:Net_decomposition}
\end{equation}
The neural network is constructed such that it can approximate the
distribution of type Eq.\ref{eq:Net_decomposition}. The inputs to the
neural network are $\alpha,\beta$ and $\gamma$ drawn uniformly at random
and the outputs are the conditional probabilities i.e. $P_{A}\left(a\vert\beta,\gamma\right),P_{B}\left(b\vert\gamma,\alpha\right)$
and $P_{C}\left(c\vert\alpha,\beta\right)$. 

\begin{figure}[t]
	\includegraphics[width=0.45\textwidth]{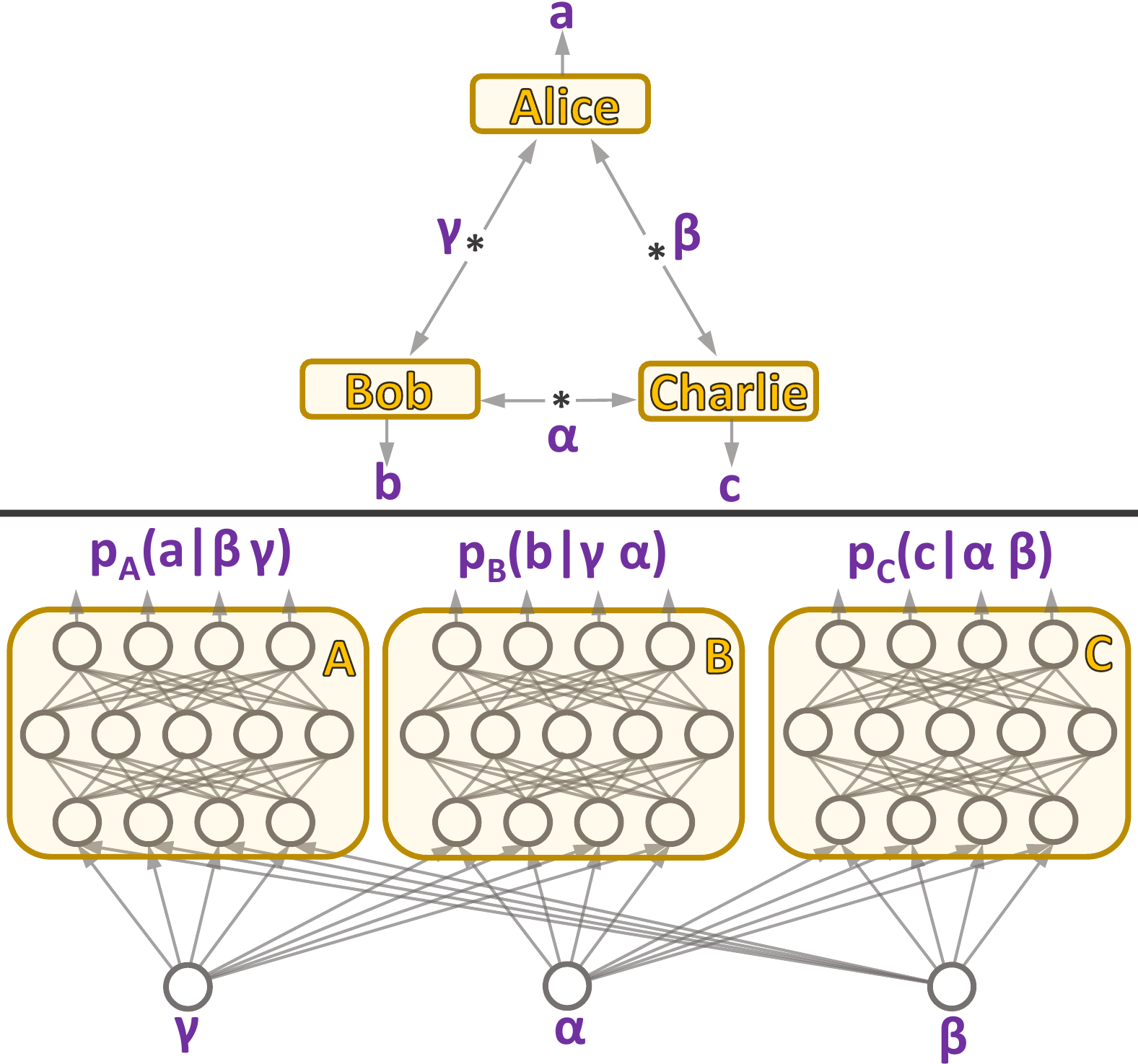}
	\caption{Reprinted with permission from Krivachy \textit{et al.}\cite{krivachy2019neural}. (Top) The triangle network configuration. (Bottom) The topology of the neural network is selected such that it reproduces distributions compatible with
the triangle configuration. } 
	\label{Fig0}
\end{figure}
The cost function  is chosen to be 
any measure of the distance between the target distribution
$P_\text{t}$ and the network's output $P_\text{M}$. The authors employ the techniques
to a few other cases, such as the elegant distribution and a distribution
proposed by Renou \textit{et al.}~\cite{renou2019genuine}. The application of the technique
to the elegant distribution suggests that the distribution is indeed
nonlocal as conjectured in \cite{gisin2019entanglement}. Furthermore, the distribution proposed by
Renou \textit{et al.} appears to have nonlocal features for some parameter regime. For the sake of completeness, now we discuss the elegant distribution and the distribution proposed by Renou \textit{et al.} 

\emph{Elegant distribution --} The distribution is generated by three parties performing entangled measurements on entangles systems. The three parties share singlets i.e. $\vert \psi^{-}\rangle = \frac{1}{\sqrt{2}}\left(\vert01\rangle - \vert10\rangle \right)$ Every party performs entangled measurements on their two qubits. The eigenstates of the entangled measurements are given by
\begin{equation}
    \vert \chi_j \rangle = \sqrt{\frac{3}{2}}\vert \alpha_j,-\alpha_j \rangle +i \frac{\sqrt{3}-1}{2} \vert \psi^{-}\rangle,
\end{equation}
where $\vert \alpha_j \rangle $ are vectors symmetrically distributed on the Bloch sphere i.e. point to the vertices of a tetrahedron, for $j \in \{1,2,3,4\}.$ 

\emph{Renou \textit{et al.} distribution --}  The distribution is generated by three parties sharing the entangled state $\vert \phi^{+} \rangle = \frac{1}{\sqrt{2}}\left(\vert00\rangle + \vert11\rangle \right)$ and performing the same measurement on each of their two qubits. The measurements are characterized by a single parameter $\kappa \in \left[\frac{1}{\sqrt{2}},1\right]$ with eigenstates $\vert 01 \rangle, \vert 10 \rangle, u\vert 00 \rangle + \sqrt{1-u^2} \vert11\rangle $ and $\sqrt{1-u^2} \vert 00 \rangle -u \vert11\rangle$.

\section{\label{sec:Oracle_Network} Machine Learning for optimizing Settings for Bell Nonlocality}
Bell inequalities have become a standard tool to reveal the non-local structure of quantum mechanics. However, finding the best strategies to violate a given Bell inequality can be a difficult task, especially for many-body settings or even non-convex scenarios. Especially the latter setting is challenging, as standard optimisation tools are unable to be applied to this case. 
To violate a given Bell inequality, two inter-dependent tasks have to be addressed: Which measurements have to be performed to reveal the non-locality? And which quantum states show the maximal violation?
Recently, Dong-Ling Deng has approached the latter task for convex settings with many-body Bell inequalities using restricted Boltzmann machines~\cite{deng2018machine}. Bharti {\em et al.} \cite{bharti2019teach} have approached both tasks in conjunction for both convex and non-convex inequalities.

\subsection{Detecting Bell nonlocality with many-body states}

Several methods from machine learning have been adopted to
tackle intricate quantum many-body problems \cite{carleo2017solving, saito2018machine,deng2018machine,gao2017efficient}. Dong-Ling Deng \cite{deng2018machine} employs machine learning techniques to detect quantum nonlocality in many-body systems using the restricted Boltzmann
machine (RBM) architecture. The key idea can be split into two parts:
\begin{itemize}
\item After choosing appropriate measurement settings, Bell inequalities for convex scenarios can be expressed as an operator. This operator can be thought of as an Hamiltonian. The eigenstate with the maximal eigenvalue is the state that gives the maximum violation for the underlying Bell inequality. This state can be found by calculating the ground state of the negative of the Hamiltonian. %
\item Finding the ground state of a quantum Hamiltonian is QMA-hard and thus in general difficult. However, using heuristic techniques involving the RBM architecture, the problem is recast into the task of finding the approximate answer in some cases.
\end{itemize}
Techniques like density matrix renormalization group (DMRG) \cite{schollwock2005density}, projected entangled pair states (PEPS) \cite{verstraete2008matrix} and multiscale entanglement renormalization ansatz (MERA) \cite{vidal2008class} are traditionally used to find (or approximate) ground states of many-body Hamiltonians. But these techniques only work reliably for optimization problems involving low-entanglement states. Moreover, DMRG works well only for systems with short-range interactions in the one-dimensional case. As evident from references \cite{carleo2017solving}, RBM can represent quantum many-body problems beyond 1-D and low-entanglement.

\subsection{\label{sec:AI_Game}Playing  Bell Nonlocal Games}
\begin{figure}[htp] 
	\includegraphics[width=0.4\textwidth]{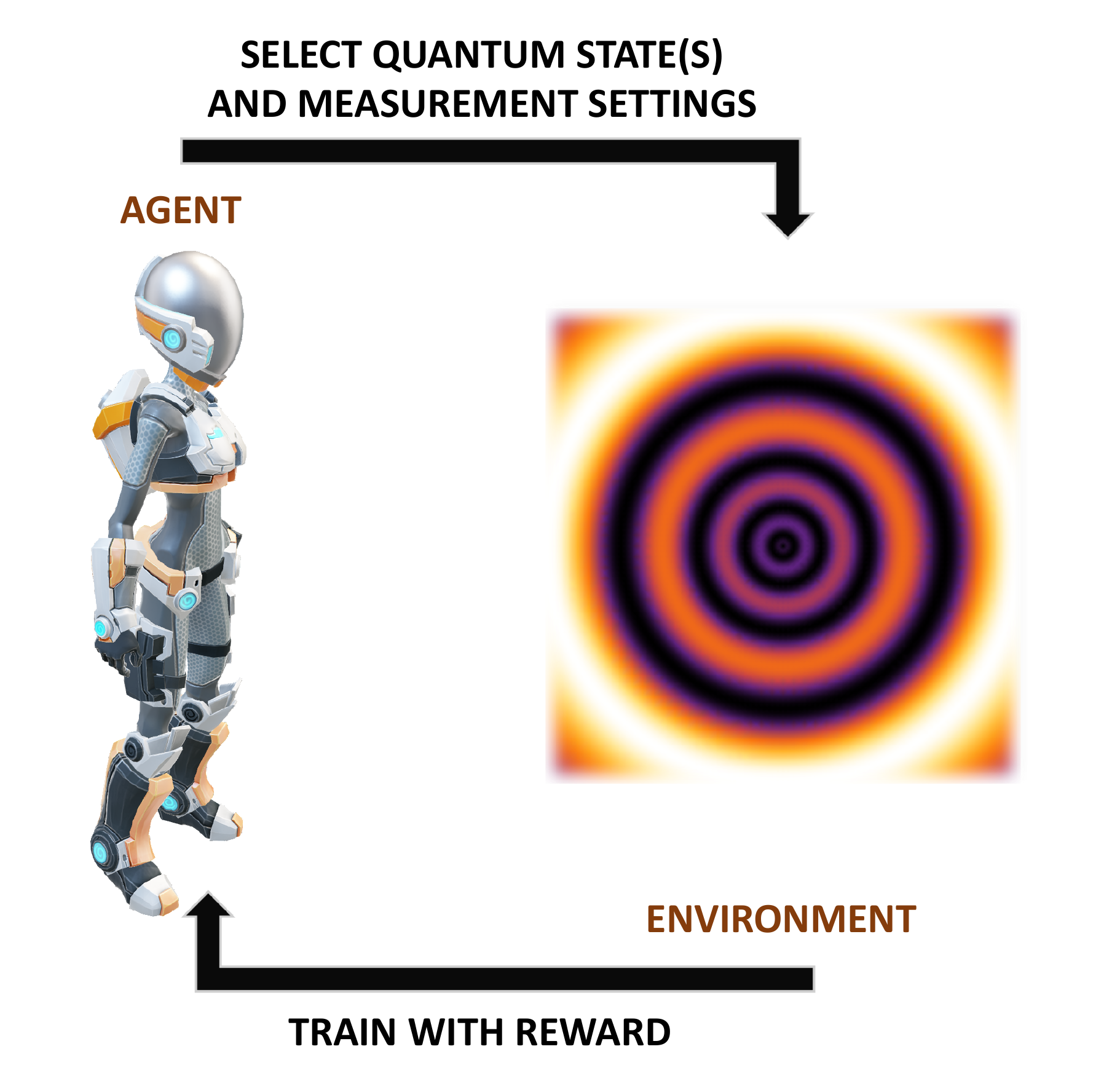}
	\caption{Reprinted with permission from Ref \cite{bharti2019teach}. Playing Bell games with AI: In \cite{bharti2019teach}, the authors train AI agent to play various Bell nonlocal games. The agent interacts with the quantum system by choosing quantum states and measurement angles, and measures resulting violation of the Bell inequalities. Over repeated games, the agent trains himself using past results to realize maximal violation of Bell inequalities.}
	\label{Fig1}
\end{figure}

Prediction of winning strategies for (classical) games and decision-making processes with reinforcement learning (RL) has made significant progress in game theory in recent years. Motivated partly by these results, the authors in Bharti {\em et al.} \cite{bharti2019teach}  have looked at a game-theoretic formulation of Bell inequalities (known as Bell games) and applied machine learning techniques to it. To violate a Bell inequality, both the quantum state as well as the measurements performed on the quantum states have to be chosen in a specific manner. The authors transform this problem into a decision making process. 
This is achieved by choosing the parameters in a Bell game in a sequential manner, e.g. the angles of the measurement operators, the angles parameterizing the quantum states, or both. 
Using RL, these sequential actions are optimized for the best configuration corresponding to the optimal/near-optimal quantum violation (see Fig.\ref{Fig1}). The authors train the RL agent with a cost function that encourages high quantum violations via proximal policy optimization -  a state-of-the-art RL algorithm that combines two neural networks. The approach succeeds for well known convex Bell inequalities, but it can also solve Bell inequalities corresponding to non-convex optimization problems, such as in larger quantum networks. 
So far, the field has struggled solving these inequalities; thus, this approach offers a novel possibility towards finding optimal (or near-optimal) configurations. 

Furthermore, the authors present an approach to find settings corresponding to maximum quantum violation of Bell inequalities on near-term quantum computers. The quantum state is parameterized by a circuit consisting of single-qubit rotations and CNOT gates acting on neighbouring qubits arranged in a linear fashion. Both the gates and the measurement angles are optimized using a variational hybrid classic-quantum algorithm, where the classical optimization is performed by RL.
The RL agent learns by first randomly trying various measurement angles and quantum states. Over the course of the training, the agent improves itself by learning from experience, and is capable of reaching the maximal quantum violation.

\section{\label{sec:MLandIneq} Machine Learning-Assisted State Classification}

A crucial problem in quantum information is identifying the degree of entanglement within a given quantum state. For Hilbert space dimensions up to $6$, one can use Peres-Horodecki
criterion, also known as positive partial transpose criterion (PPT) to distinguish entangled and separable states. 
However, there is no generic observable or entanglement witness as it is in fact a NP-hard problem \cite{horodecki2009quantum}. 
Thus, one must rely on heuristic approaches.
This poses a fundamental question: Given partial or full information about the state, are there ways to classify whether it is entangled or not? Machine learning has offered a way to find  answers to this question.

\subsection{Classification with Bell Inequalities}

In reference \cite{ma2018transforming}, the authors blend Bell inequalities with a feed-forward neural
network to use them as state classifiers. The goal is to classify
states as entangled or separable. If a state violates
a Bell inequality, it must be entangled. However, Bell inequalities
cannot be used as a reliable tool for entanglement classification i.e.
even if a state is entangled, it might not violate an entanglement witness based on the Bell inequality. For example, the density matrix $\rho=p\vert\psi_{-}\rangle\langle\psi_{-}\vert+\left(1-p\right)\frac{I}{4}$
violates the CHSH inequality only for $p>\frac{1}{\sqrt{2}}$, but
is entangled for $p>\frac{1}{3}$ \cite{werner1989quantum}. Moreover, given a Bell inequality,
the measurement settings that witness entanglement of a quantum
state (if possible) depend on the quantum state. Prompted by these
issues, the authors ask if they can transform Bell inequalities into
a reliable separable-entangled states classifier. The coefficients of the terms  in the CHSH inequalities are $(1,1,1,-1).$ The local hidden variable bound on the inequality is $2$ (see equation \ref{CHSH_in}). Assuming fixed measurement settings corresponding to CHSH inequality, the authors examine whether
it is possible to get better performance in terms of entanglement classification compared with the values $\left(1,1,1,-1,2\right)$. The main challenge to answer such a  question in a supervised learning setting is to get labelled data that is verified to be either separable or entangled. 

To train the neural network, the correlation vector corresponding to the
appropriate Bell inequality was chosen as input with the
state being entangled or separable $\left(1\text{ versus }0\right)$
as corresponding output. The correlation vector contains the expectation of the product of Alice's and Bob's measurement operators. The performance of the network improved as
the model capacity was increased, which hints that the hypothesis
which separates entangled states from separable ones must be sufficiently
``complex.'' The authors also trained a neural network to distinguish
bi-separable and bound-entangled states.

\subsection{Classification by Representing Quantum States with Restricted Boltzmann Machines}
Harney \emph{et al.} \cite{harney2019entanglement} use reinforcement learning with RBMs to detect entangled states. RBMs have demonstrated being capable of learning complex quantum states. The authors modify RBMs such that they can only represent separable states.  This is achieved by separating the RBM into $K$ partitions that are only connected within themselves, but not with the other partitions. Each partition represents a (possibly) entangled subsystems, that however is not entangled with the other partitions. This choice enforces a specific $K$-separable Ansatz of the wavefunction. 
This RBM is trained to represent a target state. If the training  converges, it must be representable by the Ansatz and thus be $K$-separable. However, if the training does not converge, the Ansatz is insufficient and the target state is either of a different $K'$-separable form or fully entangled.

\subsection{Classification with Tomographic Data}
Can tools from machine learning help to distinguish entangled and separable states given the full quantum state (e.g. obtained by quantum tomography) as an input? Two recent studies address this question.

In Lu {\em et al.} \cite{lu2018separability}, the authors detect entanglement by employing classic (i.e. non-deep learning) supervised learning. 
To simplify data generation of entangled and separable states, the authors approximate the set of separable states by a convex hull (CHA). States that lie outside the convex hull are assumed to be most likely entangled. 
For the decision process, the authors use ensemble training via bootstrap aggregating (bagging). Here, various supervised learning methods are trained on the data, and they form a committee together that decides whether a given state is entangled or not. The algorithm is trained with the quantum state and information encoding the position relative to the convex hull as inputs. The authors show that accuracy improves if bootstrapping of machine learning methods is combined with CHA.

In a different approach Goes {\em et al.}~\cite{goes2020automated}, the authors present an automated machine learning approach to classify random states of two qutrits as separable (SEP), entangled with positive partial transpose (PPTES) or entangled with negative partial transpose (NPT). 
For training, the authors elaborate a way to find enough samples to train on. The procedure is as follows: A random quantum state is sampled, then using the General Robustness of Entanglement (GR) and PPT criterion, it is classified to either SEP, PPTES or NPT. The GR measures the closeness to the set of separable states. %
The authors compare various supervised learning methods to distinguish the states. The input features fed into the machine are the components of the quantum state vector and higher-order combinations thereof, whereas the labels are the type of entanglement. 
Besides, they also train to estimate the GR with regression techniques and use it to validate the classifiers.

\section{\label{sec:Interpretations} Neural Networks as ``Hidden" Variable Models for Quantum Systems }
Understanding why deep neural networks work so well is an active area
of research. The presence of the word ``hidden'' for hidden variables
in quantum foundations and hidden neurons in deep learning neurons
may not be that accidental. Using conditional Restricted Boltzmann
machines (a variant of Restricted Boltzmann machines), Steven Weinstein
provides a completion of quantum theory in reference \cite{weinstein2018neural}.
The completion, however, doesn't contradict Bell's theorem as the assumption
of ``statistical independence'' is not respected. The statistical
independence assumption demands that the complete description of the
system before measurement must be independent of the final measurement
settings. The phenomena where apparent nonlocality is observed by
violating statistical independence assumption is known as ``nonlocality
without nonlocality'' \cite{weinstein2009nonlocality}.

In a Bell-experiment corresponding to CHSH scenario, the detector
settings $\alpha\in\left\{ a,a^{\prime}\right\}$ and $\beta\in\left\{ b,b^{\prime}\right\}$, and
the corresponding measurement outcomes $x_{\alpha}\in\left\{ 1,-1\right\} $and
$x_{\beta}\in\left\{ 1,-1\right\} $ for a single experimental trial
can be represented as a four-dimensional vector $\left\{ \alpha,\beta,x_{\alpha},x_{\beta}\right\} .$
Such a vector can be encoded in a binary vector $V=\left(v_{1},v_{2},v_{3},v_{4}\right)$
where $v_{i}\in\left\{ 0,1\right\} .$ Here, $(+1)/(-1)$ has been
mapped to $0/(+1).$ The four-dimensional binary vector $V$ represents
the value taken by four visible units of an RBM. The dependencies
between the visible units is encoded using sufficient number of hidden
units $H=\left(h_{1},h_{2},\cdots,h_{j}\right).$ With four hidden
neurons, the authors could reproduce the statistics predicted by
EPR experiment with high accuracy. For example, say the vector $V=\left(0,1,1,0\right)$
occurs in $3\%$ of the trials, then after training the machine would
associate $P\left(V\right)\approx0.03.$ Quantum mechanics gives us
only the conditional probabilities $P\left(x_{\alpha},x_{\beta}\vert\alpha,\beta\right)$
and thus learning joint probability using RBM is resource-wasteful.
The authors harness this observation by encoding the conditional statistics
only in a conditional RBM (cRBM).

The difference between a cRBM and RBM is that the units corresponding
to the conditioning variables (detector settings here) are not dynamical
variables. There are no probabilities assigned to conditioning variables, and thus the only probabilities generated by cRBM are conditional
probabilities. This provides a more compact representation compared to an RBM.

\section{\label{sec:more}A Few More Applications}
Work on quantum foundations has led to the birth of quantum computing and quantum information. Recently, the amalgamation of quantum theory and machine learning has led to a new area of research, namely quantum machine learning \cite{schuld2018supervised}. For further reading on quantum machine learning, refer to \cite{schuld2015introduction} and references therein. Further, for recent trends and exploratory works in quantum machine learning, we refer the reader to   \cite{dunjko2020non} and references therein.

Techniques from machine learning have been used to discover new quantum experiments \cite{melnikov2018active, krenn2016automated}. In reference \cite{krenn2016automated}, Krenn \textit{et. al.} provide a computer program which helps designing novel quantum experiments. The program was called Melvin. Melvin provided solutions which were quite counter-intuitive and different than something a human scientist ordinarily would come up with. Melvin's solutions have led to many novel results \cite{gu2018gouy, wang2017generation, babazadeh2017high, schlederer2016cyclic, erhard2018experimental, malik2016multi}. The ideas from Melvin have further provided machine generated proofs of Kochen-Specker theorem \cite{pavivcic2019automated}.

\section{\label{sec:Conclusion} Conclusion and Future Work}
In this survey, we discussed the various applications of machine learning
for problems in the foundations of quantum theory such as determination of the quantum bound for Bell inequalities,
the classification of different behaviors in local/nonlocal sets,
using hidden neurons as hidden variables for completion of quantum
theory, training AI for playing Bell nonlocal games, ML-assisted state
classification, and so forth. Now we discuss a few open questions at
the interface. Some of these open questions have been mentioned in
references \cite{canabarro2019machine,krivachy2019neural,bharti2019teach,deng2018machine}

Witnessing Bell nonlocality in many-body systems is an active area of research \cite{tura2014detecting,deng2018machine}.
However, designing experimental-friendly many-body Bell inequalities
is a difficult task. It would be interesting if machine learning could
help design optimal Bell inequalities for scenarios involving many-body
systems. In reference \cite{deng2018machine}, the author used RBM based
representation coupled with reinforcement learning to find near-optimal
quantum values for various Bell inequalities corresponding to various
convex Bell scenarios. It is well known that optimization becomes comparatively easier once the representation gets compact. It would be interesting if one can use other neural
networks based representations such as convolutional neural networks
for finding optimal (or near-optimal) quantum values.

As mentioned in reference \cite{canabarro2019machine}, it is an excellent idea to deploy techniques like anomaly detection for the detection of non-classical
behaviors.  This can be done by subjecting the machine to training with local behaviors only. 

In many of the applications, e.g. classification of entangled states, the computer gives a guess, but we are not sure about the correctness. This is never said and cannot be overlooked, as it is a limitation of these applications. The intriguing question is to understand how the output of the computer can be employed to provide a certification of the result, for instance, an entanglement witness. One could use ideas from probabilistic machine learning in such cases \cite{ghahramani2015probabilistic}. The probabilistic framework, which explains how to represent and handle uncertainty about models and predictions, plays a fundamental role in scientific data analysis. Harnessing the tools from probabilistic machine learning such as Bayesian optimization and automatic model discovery could be conducive in understanding how to utilize machine output to provide a certification of the result.

In reference \cite{bharti2019teach}, the authors used reinforcement learning
to train AI to play Bell nonlocal games and obtain optimal (or near-optimal)
performance. The agent is offered a reward at the end of each epoch
which is equal to the expectation value of the Bell operator corresponding
to the state and measurement settings chosen by the agent. Such a
reward scheme is sparse and hence it might not be scalable. It would
be interesting to come up with better reward schemes.
Furthermore, in this approach only a single agent tries to
learn the optimization landscape and discovers near-optimal (or optimal) measurement settings
and state. It would be exciting to extend the approach to multi-agent
setting where every space-like separated party is considered a separate agent. It is worth mentioning that distributing the actions and observations of a single agent into a list of agents reduces the dimensionality of agent inputs and outputs. Furthermore, it also dramatically improves the amount of training data produced per step of the environment. Agents learn better if they tend to interact as compared to the case of solitary learning.

Bell inequalities separate the set of local behaviours from the set of non-local behaviours. The analogous boundary separating quantum from the post-quantum set is known as quantum Bell inequality \cite{masanes2005extremal, thinh2019computing}. Finding quantum Bell inequalities is an interesting and challenging problem. However, one can aim to obtain the approximate expression by supervised learning with the quantum Bell inequalities being the boundary separating
the quantum set from the post-quantum set. Moreover,
it is interesting to see if it is possible to guess physical
principles by merely opening the neural-network black box.

Driven by the success
of machine learning in Bell nonlocality, it is genuine to ask if the
methods could be useful to solve problems in quantum steering and
contextuality. Recently, ideas from the exclusivity graph approach to contextuality
were used to investigate problems involving causal inference \cite{poderini2019exclusivity}. Ideas from quantum foundations could further assist in developing a deeper understanding of machine learning or in general artificial
intelligence. 

In artificial intelligence, one of the tests to distinguish between
humans and machines is the famous ``Turing Test (TT)'' due to Alan
Turing \cite{russell2002artificial,saygin2000turing}. The purpose of TT is to determine if a computer is linguistically
distinguishable from a human. In TT, a human and a machine are sealed
in different rooms. A human jury who does not know which room contains
a human and which room not, asks questions to them, by email, for example.  Based on the returned outcome, if the judge cannot do better than fifty-fifty, then the machine in question is said to have passed TT. The task of distinguishing the humans from machine based on the statistics of the answers (say output $a)$ given questions (say input $x)$ is a statistical distinguishability test assuming the rooms plus its inhabitants as black boxes. In the black-box approach to quantum theory, experiments are regarded as a black box where the experimentalist introduces a measurement (input) and obtains the outcome of the measurement (output). One of the central goals of this approach is to deduce statements regarding the contents of the black box based on input-output statistics \cite{acin2017black}. It would be nice to see if techniques from the black-box approach to quantum theory could be connected to TT.%

\begin{acknowledgments}
We wish to acknowledge the support of the Ministry of Education and the National Research Foundation, Singapore. We thank Valerio Scarani for valuable discussions.
\end{acknowledgments}

\section*{Data Availability}
Data sharing is not applicable to this survey as no new data were created or analyzed in this study.

\bibliography{references}
\bibliographystyle{unsrt}

\end{document}